\pdfoutput=1
\documentclass[a4paper,fleqn,usenatbib]{mnras}

\usepackage{newtxtext,newtxmath}
\usepackage[T1]{fontenc}
\usepackage{ae,aecompl}

\usepackage{graphicx}	% Including figure files
\usepackage{amsmath}	% Advanced maths commands
\usepackage{amssymb}	% Extra maths symbols
\usepackage{threeparttable}
\usepackage{tablefootnote}
\usepackage{booktabs}
\usepackage{longtable}
\usepackage{colortbl}

\newcommand{\AS}{ASASSN-14li}

\title[ASASSN-14li long-term radio and X-ray evolution]{Long-term radio and X-ray evolution of the tidal disruption event ASASSN-14li}

\author[J. S. Bright et al.]{J. S. Bright,$^{1}$\thanks{E-mail: joe.bright@physics.ox.ac.uk}
R. P. Fender,$^{1}$
S. E. Motta,$^{1}$
K. Mooley,$^{1}$
Y. C. Perrott,$^{2}$
\newauthor
S. van Velzen,$^{3}$
S. Carey,$^{2}$
J. Hickish,$^{2}$
N. Razavi-Ghods,$^{2}$
D. Titterington,$^{2}$
\newauthor
P. Scott,$^{2}$
K. Grainge,$^{4}$
A. Scaife,$^{4}$
T. Cantwell,$^{4}$
C. Rumsey$^{2}$
\\
$^{1}$Department of Physics, University of Oxford, Denys Wilkinson Building, Keble Road, Oxford OX1 3RH, UK\\
$^{2}$Astrophysics Group, Cavendish Laboratory, 19 J. J. Thomson Avenue, Cambridge CB3 0HE, UK\\
$^{3}$Department of Physics and Astronomy, The Johns Hopkins University, Baltimore, MD 21218, USA\\
$^{4}$Jodrell Bank Centre for Astrophysics, Alan Turing Building, School of Physics and Astronomy, University of Manchester, M13 9PL, UK
}

\date{Accepted XXX. Received YYY; in original form ZZZ}

\pubyear{2017}

\begin{document}
\label{firstpage}
\pagerange{\pageref{firstpage}--\pageref{lastpage}}
\maketitle

\begin{abstract}
%This is a simple template for authors to write new MNRAS papers.
%The abstract should briefly describe the aims, methods, and main results of the paper.
%It should be a single paragraph not more than 250 words (200 words for Letters).
%No references should appear in the abstract.
We report on late time radio and X-ray observations of the tidal disruption event candidate ASASSN-14li, covering the first 1000 days of the decay phase. For the first $\sim200$ days the radio and X-ray emission fade in concert. This phase is better fit by an exponential decay at X-ray wavelengths, while the radio emission is well described by either an exponential or the canonical $t^{-5/3}$ decay assumed for tidal disruption events. The correlation between radio and X-ray emission during this period can be fit as $L_{R}\propto L_{X}^{1.9\pm0.2}$. After 400 days the radio emission at $15.5\,\textrm{GHz}$ has reached a plateau level of $244\pm8\,\mu\textrm{Jy}$ which it maintains for at least the next 600 days, while the X-ray emission continues to fade exponentially. This steady level of radio emission is likely due to relic radio lobes from the weak AGN-like activity implied by historical radio observations. We note that while most existing models are based upon the evolution of ejecta which are decoupled from the central black hole, the radio : X-ray correlation during the declining phase is also consistent with core jet emission coupled to a radiatively efficient accretion flow.
\end{abstract}


%%%%%%%%%%%%%%%%%%%%%%%%%%%%%%%%%%%%%%%%%%%%%%%%%%

%%%%%%%%%%%%%%%%% BODY OF PAPER %%%%%%%%%%%%%%%%%%

\section{Introduction}
A tidal disruption event (TDE) occurs when a stellar object passes close to a supermassive black hole (SMBH) on a highly eccentric orbit which is potentially misaligned with the black hole's spin, and is torn apart when strong tidal forces overcome the star's self gravity. Approximately half of the mass of the disrupted object forms a complex accretion flow onto the SMBH, while the other half becomes unbound and is ejected from the system (see e.g. \citealt{1988Natur.333..523R} and \citealt{2014ApJ...783...23G}). TDEs can be identified through their strong blue continuum, showing significant brightening in optical, UV -- and sometimes X-ray -- compared to archival observations of the host galaxy. Spectral signatures of TDEs include broad hydrogen (Balmers series) emission lines as well as emission in \ion{He}{i} and \ion{He}{ii}, although the abundance of specific elements varies significantly between events (\citealt{2014ApJ...793...38A}). The critical radius for disruption, where tidal effects become dominant, is $R_{p}\sim R_{\textrm{tidal}}=R_{\star}(M_{\textrm{BH}}/M_{\star})^{1/3}$, where $R_{p}$ is the pericentre distance of the star's orbit. The tidal radius must be outside the black hole's Schwarzschild radius in order for the flaring associated with these events to be observed.\\

Tidal disruption event candidates are typically discovered through optical (e.g. \citealt{vanvelzenOPTTDE2011}; \citealt{holoien14li2016}), UV (e.g. \citealt{gezariUVTDE2009}), or X-ray emission (e.g. \citealt{komossa1999}; \citealt{esquej2007}) from the initial accretion flow and resulting flare in transient surveys, and are confirmed as TDEs based on their decay characteristics and spectra. The accretion flow formed in a TDE is unlike classical AGN accretion disks, consisting of orders of magnitude less mass but producing significantly higher accretion rates, which can be (initially) super-Eddington. The material that remains bound to the SMBH is expected to form an accretion disk, with the circularisation timescale depending on relativistic precession effects, which dictate the chance of stream self-interaction (\citealt{2013MNRAS.434..909H}; \citealt{2016MNRAS.455.2253B}; \citealt{0004-637X-809-2-166}) and are in part dictated by the spin of the SMBH. It has also been shown (\citealt{2013MNRAS.434..909H}) that the radiative cooling efficiency can affect the circularisation timescale, as well as the thickness of the disk. The observational signatures of such an accretion flow will depend on the circularisation, radiative cooling and viscous timescales which are described and discussed by \citet{1989ApJ...346L..13E} and \citet{2016MNRAS.455.2253B}. Alternatively, it has been suggested (\citealt{piranSELF2015}; \citealt{2017ApJ...837L..30P}) that the dominant component of the optical/UV emission could be the result of the bound debris stream self interacting (and shocking) as it circles the black hole, with the disk predominantly contributing X-ray emission.\\

\AS\ was discovered by the All Sky Automated Search for Supernova (ASASSN) on UT 2014--11--22.63 (MJD 56983.6) as a 16.5 magnitude source in the V-band (\citealt{2014ATel.6777....1J}). The position of the source was found to be consistent with the centre of the post-starburst galaxy PGC 043234, with a measured projected separation of 0.04\arcsec. This galaxy is at redshift $z=0.0206$ with a luminosity distance of $90.3\,\textrm{Mpc}$ (for cosmological parameters $H_{0}=73\,\textrm{km}\textrm{s}^{-1}\,\textrm{Mpc}^{-1}$, $\Omega_{\textrm{matter}}=0.27$ and $\Omega_{\Lambda}=0.73$). It was established through archival X-ray observations of PGC 043234 from the ROSAT All-Sky Survey (\citealt{1999A&A...349..389V}) that the galaxy does not contain an efficiently accreting AGN, with the count rate implying a luminosity orders of magnitude below standard active nuclei (e.g. \citealt{2015Natur.526..542M}). A small number (currently six) of confirmed TDEs, including \AS,\ have also been detected at radio wavelengths and the population may form a bi-modal distribution, consisting of more common non-relativistic `thermal' events and rarer relativistic jets. Three events (Swift J1644+57; \citealt{2011Natur.476..421B}, \citealt{2011Natur.476..425Z}, Swift J2058+05; \citealt{2012ApJ...753...77C}, Swift J1112.2; \citealt{gbrown2017}) have isotropic $\sim5\,\textrm{GHz}$ luminosities of between $10^{40}$ and $10^{42}\,\textrm{erg}\:\textrm{s}^{-1}$ whereas the rest (IGR J12580+0134; \citealt{2015ApJ...809..172I}, XMMSL1 J0740-85; \citealt{2017ApJ...837..153}, \AS;\ \citealt{2016Sci...351...62V}, \citealt{2016ApJ...819L..25A}) have luminosities in the range $10^{37}$ to $10^{39}\,\textrm{erg}\:\textrm{s}^{-1}$ at similar frequencies. The higher power events are believed to result from observing down the axis of a relativistic jet, resulting in the energy of photons being significantly boosted. Even accounting for boosting, these relativistic events have a higher total energy output than their thermal counterparts. The origin of the radio emission from the thermal events is currently uncertain, with transient jets (\citealt{2016Sci...351...62V}), non-relativistic winds (\citealt{2016ApJ...819L..25A}) and shocks driven by unbound material (\citealt{2016ApJ...827..127K}) all feasible scenarios. \AS\ is by far the best studied of the `thermal' TDE category, having been observed extensively at Optical, UV, X-ray (where \AS\ is unusually loud for an optically selected TDE) and radio wavelengths. The high cadence X-ray and radio observations in particular allow for the X-ray/radio coupling to be probed.\\

We organise this paper as follows: in section \ref{sec:obs} we describe our radio and X-ray observations of \AS\ and their analysis. In sections \ref{sec:res} and \ref{sec:dis} we show our correlation of the multi-wavelength data, and demonstrate that \AS\ has now faded below its host galaxy's background radio emission. Our results are summarised in section \ref{sec:sum}, with all radio and X-ray data used given in the appendix.

\section{Observations}\label{sec:obs}

\subsection{\textit{Swift} X-ray observations}
\textit{Swift} (\citealt{gehrelsSWIFT2004}) observations of \AS\ were first initiated on 56991.5 and a total of 102 observing segments have been conducted to date (100 of which had photon counting (PC) mode data), with exposure times ranging between $90$ and $4000$ seconds. The 6 segments between 57819.6 and 57833.5 were triggered due to a target of opportunity request submitted by the authors, whereas all other observations are archival. The position of the source was measured as J2000 RA/Dec = $12\textrm{h}\,48\textrm{m}\,15.05\textrm{s}/$$+17\degr\ 46\arcmin\ 31.5\arcsec$ with a circular 90\% confidence region of radius 3.5\arcsec (\citealt{2007A&A...476.1401G}; \citealt{2009MNRAS.397.1177E}). The source was observed multiple times per week for the first $\sim 200$ days post flare, and more sporadically afterwards. Observations utilised the X-ray telescope (XRT) instrument in PC mode as well as the ultra violet and optical telescope (UVOT), with the exact filter depending on the observation. We are primarily interested in the X-ray data, and the \textit{Swift} XRT product generator online reduction pipeline (\citealt{2007A&A...469..379E}; \citealt{2009MNRAS.397.1177E}) was used to extract count rates in the $0.3$--$10\,\textrm{keV}$ energy band from the observations. In order to probe the X-ray : Radio luminosity correlation we converted our count rates into fluxes. To do this, we first binned the X-ray light curve of \AS\ into 6 broad time bins and extracted a spectrum from each bin. We defined the bins in a way that allowed us to both obtain acceptable S/N spectra, and to evenly sample the decay phase of \AS\.
These spectra were then fit with an absorbed black-body model, from which the flux could be calculated. We fitted the spectra with an absorbed \footnote{$N_\textrm{H} = 1.6 \times 10^{20} cm^{2}$} black-body component. The resulting best-fits show a black-body temperature variation consistent with those reported by \cite{2015Natur.526..542M}.
The relationship between flux and counts across the 6 bins was described by a linear fit, which was then used to convert the full light curve from count rate to flux. The \textit{Swift} X-ray light curve is shown in the bottom panel of figure \ref{fig:amilc}.

\subsection{Arcminute Microkelvin Imager Large Array radio observations -- Light curve}\label{sec:amiobs}
\begin{figure*}
	\includegraphics[width=2\columnwidth]{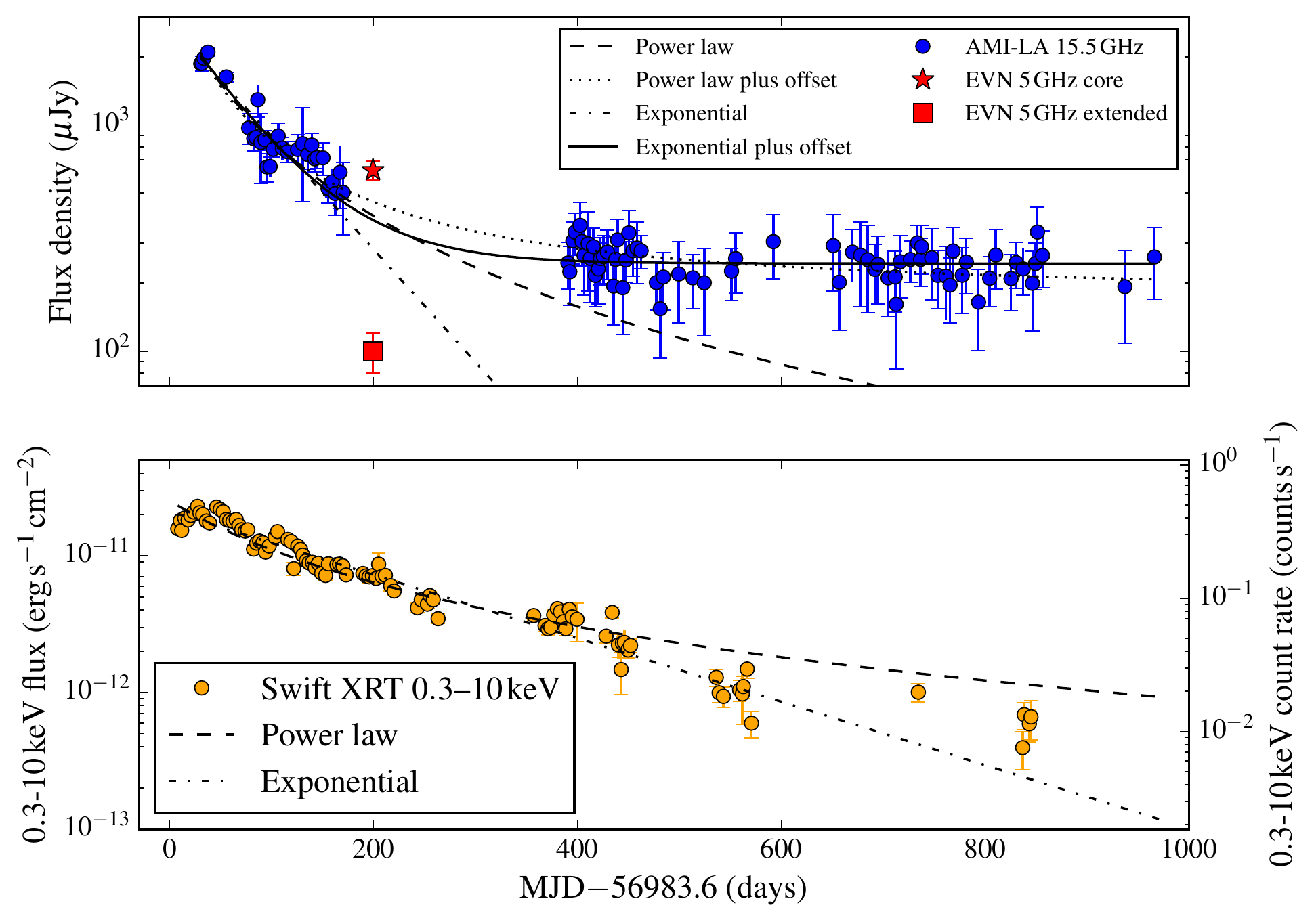}
	\caption{Radio and X-ray emission over the first 1000 days post-detection of TDE ASASSN-14li. 
{\em Upper panel:} Blue circles indicate observed radio flux measured by the AMI-LA at $15.5\,\textrm{GHz}$ ($15.7\,\textrm{GHz}$ for epochs up to $\textrm{MJD}-56983.6=200$). {\em Lower panel:} \textit{Swift} XRT observations in the $0.3$--$10\,\textrm{keV}$ energy band, measured in both counts per second and flux. In both panels error bars indicate $1$--$\sigma$ uncertainties on measurements, dashed lines show a power law decay with the exponent fixed to $t^{-5/3}$, dash-dot lines show an exponential decay. Exponential and power law decays plus a constant offset are fit to the radio data only and are shown by solid and dotted lines respectively. The red star and square indicate measurements of core and extended components (respectively) of \AS\ at $5\,\textrm{GHz}$ with the EVN (\citealt{2016ApJ...832L..10R}). Fit parameters are summarised in table \ref{tab:fits}. Note the logarithmic scale on the $y$-axis.}
	\label{fig:amilc}
\end{figure*}

Radio observations of \AS\ were initiated with the Arcminute Microkelvin Imager Large Array (\citealt{2008MNRAS.391.1545Z}; \citealt{2017arXiv170704237H}), hereafter AMI-LA, on MJD 57014.1, about 22 days after the first \textit{Swift} observations of the source. \AS\ was monitored on an approximately weekly basis (apart from a significant gap between $\sim180$ and $\sim400$ days post flare due to technical work being performed on the AMI-LA) with typical exposures lasting between 2 and 4 hours, yielding an RMS flux in a typical image of $\sim35$ and $25\,\mu$Jy, respectively. Observations pre $\textrm{MJD} - 56983.6 = 200$ were taken with an analogue correlator at an effective central frequency of 15.7\,GHz, and are published in \citet{2016Sci...351...62V}. Observations after the gap utilised the new digital correlator, which has 4096 channels across a $5\,\textrm{GHz}$ bandwidth between $13$ and $18\,\textrm{GHz}$ and are first published in this manuscript. J1255+1817 was used as the interleaved phase calibrator and was observed for $\sim2$ minutes for every $\sim10$ minutes on source while 3C286 and 3C48 were used for absolute flux calibration. The AMI-LA data were binned into 8 channels, each of width $0.625\,\textrm{GHz}$, and were calibrated and flagged for radio frequency interference (RFI) using the automated AMI reduction software pipeline \texttt{reduce} (e.g. \citealt{davies2009}). The data were then imported into CASA (\citealt{2007ASPC..376..127M}) where additional RFI flagging was performed using the \texttt{flagdata} task in \texttt{rflag} mode, which removed interference localised in time and frequency at the 3-$\sigma$ level. The data were then cleaned using the \texttt{clean} task, with a stopping threshold of 3 times the background RMS of the image and a gain of 0.1. To calculate flux measurements from \AS\ we used the python based \texttt{PySE} source extraction software, which was developed as part of the LOFAR Transient Pipeline (\citealt{2015A&C....11...25S}). A detection threshold of 3.5-sigma was used for source identification, and a 3-sigma threshold was used for fitting the source in the image plane for each observation. A 2D Gaussian, with the same dimensions as the clean beam, was used to fit the unresolved source at phase centre (corresponding to \AS). Typical dimensions of the clean beam major and minor axis FWHM are $\sim60$ and $\sim30\arcsec$ respectively. The AMI-LA light curve is shown in the top panel of figure \ref{fig:amilc}. Stacking the plateau phase radio observations in each of eight equally-spaced frequency bands, we are able to estimate a spectral index of $-0.9\pm0.5$.

%%%%%%%%%%%%%%%%%%%%% RESULTS %%%%%%%%%%%%%%%%%%%%%%%

\section{Results}\label{sec:res}

\subsection{Characterising the decays}
Radio and X-ray light curves of \AS\ for $\sim1000$ days post discovery are shown in figure \ref{fig:amilc}, along with power law and exponential fits to both data sets. The fit parameters are summarised in table \ref{tab:fits}. 
We find that the X-ray light curve is better characterised by an exponential decay, rather than the canonical (e.g. \citealt{phinney1989}) $-5/3$ exponent power law decay, demonstrated by the significantly improved reduced chi-squared statistic (table \ref{tab:fits}). The large reduced chi-squared values, however, indicate that while we may be describing the long term trends in the light curve, there is significant deviation from this simple decay model on shorter timescales. For ease of discussion we will refer to epochs up to $\textrm{MJD}-56983.6=200$ as the decay phase (although it is only the radio light curve that ceases to decay after this date), and later observations as the plateau phase. When fitting the radio light curve, we opt to consider two cases. First, we fit the decay phase during the first $\sim200$ days and find that the $-5/3$ power law explains the data marginally better than an exponential. The whole set of radio observations (decay and plateau) is well characterised by an exponential decay plus a constant offset (a -5/3 power law plus offset provided a worse fit). Whilst it would be natural to attempt to fit the decays with a varying index and start date we are unable to resolve the inherent degeneracy in these two parameters given that we do not observe the peak of the light curve. Using the most recent date of non-detection of \AS\ at optical wavelengths presented in \citet{2017MNRAS.466.4904B} as an estimate of the start time allows for a crude estimate of the power law index to be calculated. The index, when adopting this start date, is similar to -5/3 but the reduced chi-squared is still significantly worse than for an exponential decay. 

\begin{table*}
\centering
\caption{Summary of light-curve fit parameters for the exponential, exponential plus offset and power law fits shown in figure \ref{fig:amilc}. The (entire) radio light curve is best fit by an exponential plus constant offset which is used to estimate the plateau flux. The reduced chi-squared statistic, $\chi^{2}_{\textrm{red}}$, indicates that the X-ray light curve is significantly better fit by an exponential rather than a $-5/3$ power law. The large $\chi^{2}_{\textrm{red}}$ values result from deviations from simple decay models.}
\begin{threeparttable}
\label{tab:fits}
\begin{tabular}{lccccc}
    \hline
    \hline
    \multicolumn{6}{c}{Radio}\\
    Fit type & $t_{0}$ ($\textrm{MJD} - t_{\textrm{d}}$) & Decay timescale (days) & Power law index & Offset ($\mu$Jy) & $\chi^{2}_{\textrm{red}}$\\ \hline
    Exponential\tnote{a} & -- & $87\pm6$ & -- & -- & $3.24$\\
    Exponential and offset\tnote{b} & -- & $66\pm3$ & -- & $244\pm8$ & $0.99$\\
    Power law\tnote{a,c} & $-71\pm8$ & -- & $-5/3$ & -- & $2.48$\\
    Power law and offset\tnote{b,c} & $-47\pm5$ & -- & $-5/3$ & $180\pm10$ & $0.94$\\
    %Power law\tnote{d} & $-30\pm50$ & -- & $-1.2\pm0.6$ & -- & $2.51$\\
    %Power law and offset\tnote{d} & $-130\pm80$ & -- & $-3\pm1$ & $220\pm20$ & $0.91$\\ 
    \hline
    \multicolumn{6}{c}{X-ray}\\
    Fit type & $t_{0}$ ($\textrm{MJD} - t_{\textrm{d}}$) & Decay timescale (days) & Power law index & -- & $\chi^{2}_{\textrm{red}}$\\ \hline
    Exponential & -- & $187\pm6$ & -- & -- & $8.54$\\ 
    Power law\tnote{c} & $-155\pm9$ & -- & $-5/3$ & -- & $13.99$\\
    Power law\tnote{d} & $-132.35$ & -- & $-1.60\pm0.05$ & -- & $14.95$\\ \hline
\end{tabular}
\begin{tablenotes}
\item [a] Only radio measurements in decay phase used for fit.
\item [b] All radio measurements used for fit.
\item [c] Power law index fixed to $-5/3$.
\item [d] Start date fixed.
\end{tablenotes}
\end{threeparttable}
\end{table*}

\subsection{Decay Phase X-ray : Radio correlation}
\begin{figure}
	\includegraphics[width=\columnwidth]{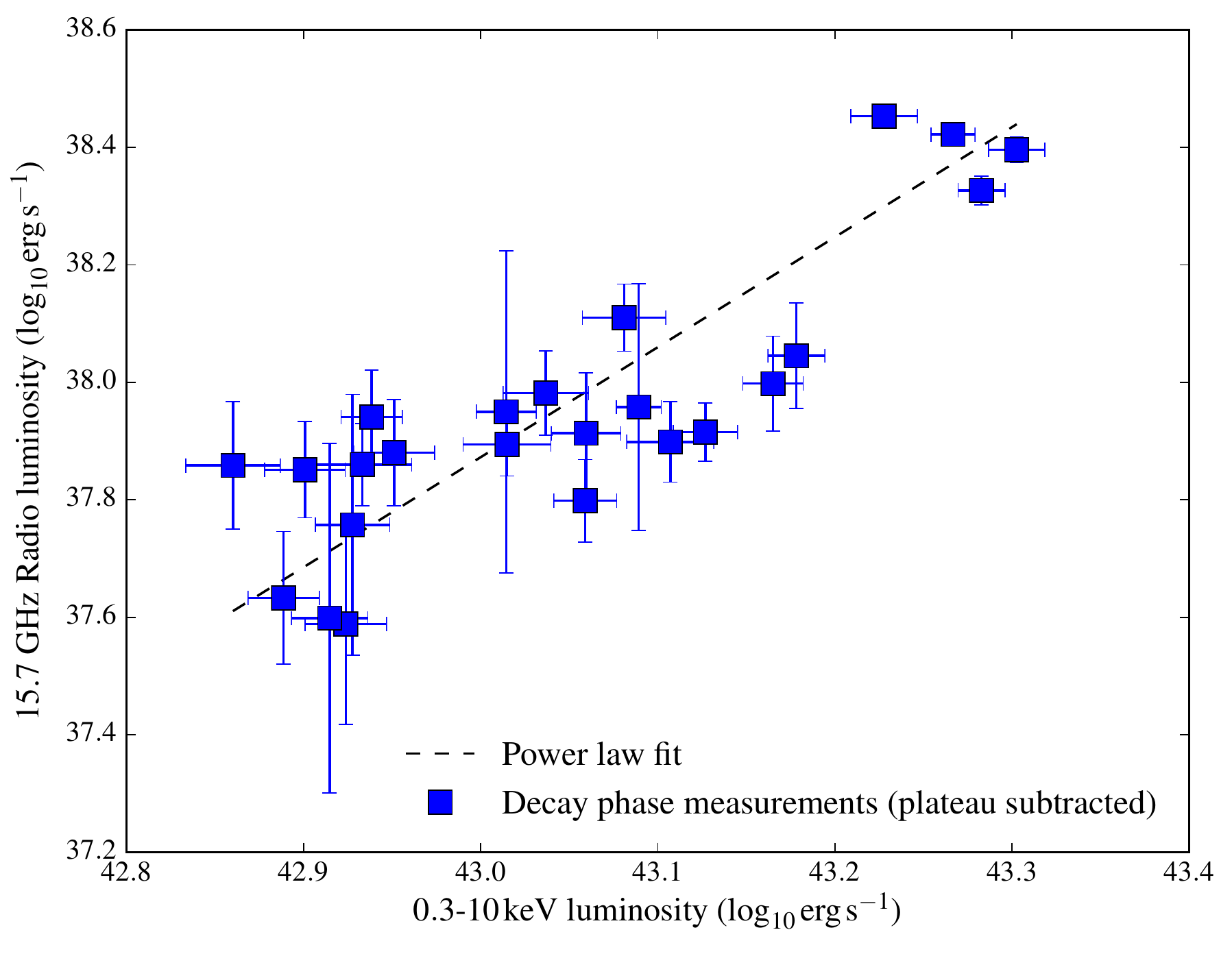}
	\caption{Radio : X-ray correlation for \AS, produced by binning the respective light curves during the decay phase and subtracting the plateau flux of $244\,\mu\textrm{Jy}$. The black dotted line is a fit to the data, with a best fit power law index of $\alpha=1.9\pm0.2$ ($L_{R}\propto L_{X}^{\alpha}$).}
	\label{fig:corr}
\end{figure}

In order to correlate the X-ray and Radio observations of \AS, which were not simultaneous, we binned each light curve into bins of width 4 days and compared the radio and X-ray flux where both had been measured in the same bin. This width was found to optimise the number of data points available for comparison. If multiple measurements at the same wavelength were contained within a specific bin then a numerical average of the data was used to represent the flux in that bin. In the case of either light curve having no data within a bin, no comparison was made. In order to ensure we were only comparing radio and X-ray emission resulting from the TDE, and not just from the host, we exclude data taken during the plateau phase from our correlation analysis. Before this date both the X-ray and radio emission were still declining and thus dominated by the TDE. We find that the radio flux density and X-ray flux (and thus the luminosities) are strongly correlated during the decay phase. The form of the relationship is $\textrm{L}_{\textrm{R}}\propto{L}_{\textrm{X}}^{1.9\pm0.2}$ and the Spearman rank correlation coefficient is 0.86. To obtain these results we subtract the plateau radio measurement of $244\pm8\,\mu\textrm{Jy}$ from the decay phase radio observations. $\textrm{L}_{\textrm{R}}$ is the isotropic, monochromatic, luminosity at $15.7\,\textrm{GHz}$ and $\textrm{L}_{\textrm{X}}$ is the X-ray luminosity in the the $0.3$--$10\,\textrm{keV}$ energy band. The correlation can be seen in figure \ref{fig:corr}.

\section{Discussion}\label{sec:dis}
\subsection{Explaining the radio plateau}
It is clear from figure \ref{fig:amilc} that the radio light curve plateaus from an initial fading phase into a  phase of constant flux of around $250\,\mu\textrm{Jy}$ which it has maintained for around 600 days. During the radio plateau period, the X-ray count rate continues to decline, showing no sign of mirroring the behaviour at $15.5\,\textrm{GHz}$. We discuss a number of possible contributors to the radio flux density plateau observed by the AMI-LA.

\subsubsection{Host galaxy radio emission}\label{sub:one}
We assume here that the plateau emission is unconnected to the TDE, which presumably continues to decline, and discuss briefly below its possible origin. The host galaxy of \AS, PGC 043234, has archival measurements in the FIRST (November 1999) and NVSS (December 1993) surveys, both at 1.4 GHz (\citealt{1995ApJ...450..559B}; \citealt{1998AJ....115.1693C}). The measured flux densities are $2.96\pm0.15$ and $3.2\pm0.4\,\textrm{mJy}$, respectively. Due to the resolution of the AMI-LA we are unable to resolve individual sources in the host galaxy and so all radio emitting elements from the galaxy will be observed to be combined into a single unresolved source. 

Firstly, supernova remnants in the host galaxy could be responsible for background emission. The small scale radio morphology of PGC 043234 has not been probed with long baseline observations (outside of the central few parsecs; \citealt{2016ApJ...832L..10R}) so we must turn to analogous galaxies which have been studied more extensively. Arp220 is a nearby ($77\,\textrm{Mpc}$) ultra-luminous infrared galaxy, which has been well studied at 18, 13, 6, 3.6 (\citealt{1998ApJ...493L..17S}; \citealt{2007ApJ...659..314P}) and most usefully $2\,\textrm{cm}$ ($15\,\textrm{GHz}$) with VLBI (\citealt{2011ApJ...740...95B}). These observations revealed a population of compact (sub-parsec) sources which are believed to be a mixture of supernovae (SNe) and supernova remnants (SNR). The $15\,\textrm{GHz}$ radio flux from these individual sources ranges from $91$ to $693\,\mu\textrm{Jy}$ which would correspond to $66$ and $504\,\mu\textrm{Jy}$ at the distance of \AS. Although this galaxy is a more extreme environment than the host of \AS,\ it illustrates that even a single radio loud SNe or SNR, or a population of fainter objects, could be responsible for the observed radio plateau of \AS. 

An alternative, and more likely, possibility is emission from relic radio lobes from past AGN activity (discussed briefly also in \citealt{2016Sci...351...62V}). Such lobes would be expected to remain constant in flux over long timescales and to have an optically thin spectrum. Considering the lack of star formation indicators (discussed in \citealt{2041-8205-830-2-L32}, \citealt{2016ApJ...819L..25A}, \citealt{2016Sci...351...62V}) in PGC 043234, the clear presence of a supermassive black hole (hence the TDE) and the EVN observations revealing the majority of the flux to be combined within a few parsecs of the central object, we conclude that past AGN activity is the more likely scenario for the quiescent component.\\

Optically thin relic AGN lobes with a spectral index of approximately $-0.9$ (consistent with both our in-band measurements and those reported towards the end of the decay phase by \citealt{2016ApJ...819L..25A}) and a $15.5\,\textrm{GHz}$ flux density of $244\,\mu\textrm{Jy}$ would correspond to a $1.4\,\textrm{GHz}$ flux density of $\sim2.1\,\textrm{mJy}$, broadly consistent the quiescent component assumed in \citet{2016ApJ...819L..25A}. It has been proposed (\citealt{2016Sci...351...62V}) that the onset of the TDE suppressed optically thick emission from a steady AGN jet in PGC 043234. Our discovery of a plateau is not inconsistent with this idea, if the system contained both AGN lobes and a compact jet pre-TDE (the sum of which gave the archival $1.4\,\textrm{GHz}$ measurements $\sim3\,\textrm{mJy}$). It is also possible (\citealt{2016ApJ...819L..25A}) that the observed decline in $1.4\,\textrm{GHz}$ emission in the 16 years since the FIRST measurement is simply due to long term AGN variability of a compact component.

\subsubsection{Radio emission still from \AS}\label{sub:two}
We also consider the possibility that the radio emission we are observing is now steady emission from the TDE. It is, however, hard to reconcile the combined observed X-ray and radio behaviour with any of the TDE models currently being considered. If the radio emission results from some form of outflow (radiatively driven wind, unbound material, discrete jet launch) we can see no reason why the radio emission would stop declining, despite it not being coupled to the likely source of X-ray emission (the accretion disk). In the case of core jet activity we would expect the X-ray and radio emission to remain coupled. In all scenarios, barring an unexpected increase in accretion rate, the decaying nature of the X-ray light curve is expected. Given the numerous possibilities for constant (on the timescale of the decay of \AS) sources of radio emission in PGC 043234 (as discussed in section \ref{sub:one}) and the fact that we can see no physical motivation for the radio emission in any of the currently considered models (core-jet emission, relativistic jet, radiatively driven wind, shocking disrupted material) to plateau, we strongly disfavour this scenario, instead attributing the quiescent radio flux density to background emission from PGC 043234.

\subsection{X-Ray/Radio correlation}
It is clear that prior to the radio plateau the radio and X-ray emission are fading together (figure \ref{fig:amilc}).
When correlating the decay phase data we subtract the plateau emission and find a strong positive correlation between the X-ray and radio flux (figure \ref{fig:corr}). The correlation can be described by a power law, with a power law index of $\alpha=1.9\pm0.2$ ($L_{R}\propto L_{X}^{\alpha}$). This correlation can be interpreted in two ways:

The more conventional explanation is that the core X-ray emission arises in the accretion flow while the radio emission is from decoupled ejecta, and both are fading independently. Two variants of this scenario are outlined in \citet{2016ApJ...827..127K} and \citet{2016ApJ...819L..25A}. Even under the assumption that one of these models is correct, it is not clear if or how our new radio and X-ray data could discriminate between them.

Alternatively, the core X-ray and radio emission may be correlated because they are coupled via an accretion flow feeding a core jet. Such an interpretation is not generally invoked for TDEs, although for this source it is also raised to explain shorter-time radio and X-ray correlations (\citealt{pasham2017}). The observed correlation, $L_{R}\propto L_{X}^{1.9\pm0.2}$, is similar to that observed for stellar mass black holes in `radio quiet' black hole X-ray binaries (e.g. \citealt{coriat2011}). While in such systems the radio spectrum is usually flatter than is observed in ASASSN-14li (Alexander et al. 2016 and this work), it is not inconsistent with the range observed (Espinasse \& Fender 2017). Typically core jet emission in stellar mass black hole binaries is seen to coincide with a hard X-ray spectrum, peaking at $\sim 100\,\mathrm{keV}$. The majority of X-ray photons from \AS\ are below $\sim 1\,\mathrm{keV}$, however not enough is known about the nature of the accretion flows in TDEs for this to exclude core jet emission. We also note that there are examples of galactic X-ray binaries (e.g. GRS 1915+105; \citealt{fender2004}) showing core radio emission with a soft X-ray spectrum, when averaged over a certain time period and at a high accretion rate.

Applying the approximate relationship between core radio luminosity and jet kinetic power presented in \citet{merloni2007} to the peak radio luminosity of \AS\ gives $L_{jet}\sim10^{43}\,\textrm{erg}\,\textrm{s}^{-1}$.
It is reassuring to note that \citet{pasham2017} find that \AS\ falls on the fundamental plane of black hole activity (\citealt{merloni2003}), although with the caveat that the X-ray luminosity was derived using the $0.3$--$1\,\mathrm{keV}$ energy band rather than the traditional $2$--$10\,\mathrm{keV}$. Further VLBI observations along the lines of those reported in \citealt{2016ApJ...832L..10R} may help to resolve uncertainties in the origin of the radio emission.

\subsection{Light curve fits}
It is clear that the forced $t^{-5/3}$ power law fit does not well describe the X-ray data compared to the exponential decay, demonstrated by the significantly increased reduced chi-squared statistic and the overestimate of the late time flux. It is, however, likely that the fit would be significantly improved if there was a better constraint on the start date ($t_{0}$) of the flare (see more estimates in \citealt{2017MNRAS.466.4904B}). In either case it is evident that a simple decay model for the X-ray emission from \AS\ is not explaining the significant short term variability in the light curve, demonstrated by the large reduced chi-square value. During the first $\sim150$ days of X-ray observations, deviations from the simple decay models are dominated by an apparently quasi-periodic signal. We do not speculate on the physical origin of this borderline significant feature, but plan to present analysis in a future work.\\

The early time radio light curve is well characterised by both a $-5/3$ power law and an exponential (although marginally better by a power law), and the entire data set is well fit by an exponential plus constant offset. The exponential fits for the light curves provide characteristic decay times for the X-ray and radio emission which are $\sim190$ and $\sim90$ days, respectively. Fitting the radio light curve with a free index suffers from the same issues as the X-ray light curve, however see \citet{auchettl2017} for a more detailed analysis of TDE decays at X-ray wavelengths.\\

If core jet coupling is present in \AS\ we might expect short term variation in the X-ray light curve to be mirrored in the radio light curve, a possibility explored in \citet{pasham2017}. The observation that the short term radio variability appears to be less pronounced than in the X-ray light curve could be explained by the fact that the radio emitting region is both larger than, and removed from, the X-ray emitting region, which would result in short term variability being washed out.\\

\begin{figure}
	\includegraphics[width=\columnwidth]{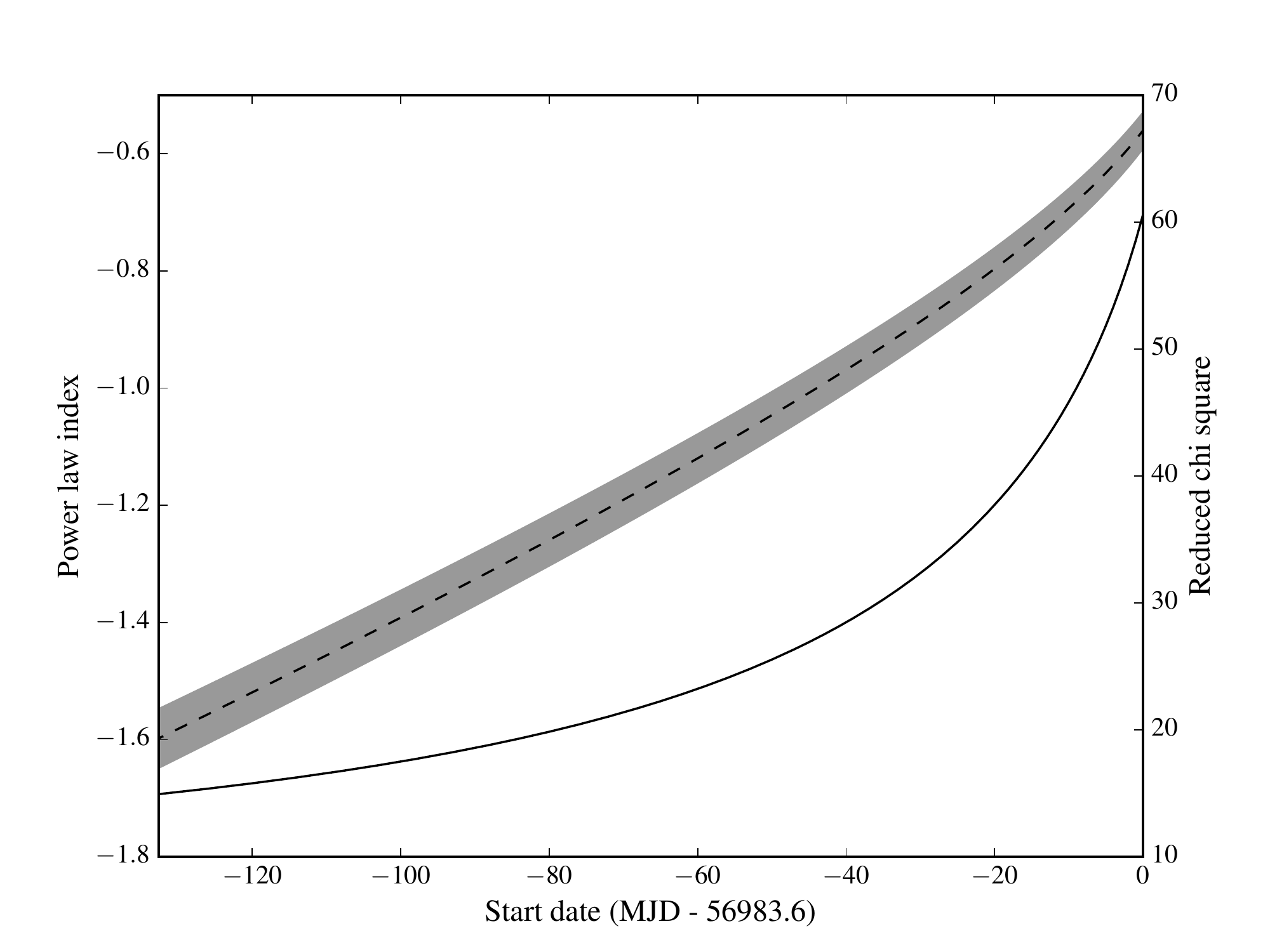}
	\caption{Best fitting power law index for the X-ray light curve (black dashed line and shaded one sigma confidence region), and the associated reduced chi-square (solid black line) as a function of fixed start date ($t_0$).}
	\label{fig:decay}
\end{figure}

The assumed start date of X-ray emission ($t_{0}$ in table \ref{tab:fits}) has a significant impact on the power law index derived to describe the decay. It is important to constrain this parameter as well as possible, as the decay index provides important information on e.g. the mass return and accretion rate and potentially the make-up of the disrupted body (e.g. \citealt{2016MNRAS.455.2253B}; \citealt{2009MNRAS.392..332L}; \citealt{0004-637X-809-2-166}). It may also be the case that these decays are not power laws at all. To investigate this effect, we fit a power law decay to the X-ray light curve with a varying $t_{0}$, beginning from the latest ASASSN non-detection ($t_0=-132.35$; \citealt{2017MNRAS.466.4904B}; although note that this is an optical, not X-ray, non-detection) and ending at the ASASSN discovery date ($t_0=0$) which is shown in figure \ref{fig:decay}. For the range of start dates considered the best fit power law index varies between $\sim-1.6$ and $\sim-0.6$ with the goodness of fit reducing dramatically with later start dates. Care must be taken, however, as earlier start dates will naturally provide a better fit and so without prompt observations that sample the rise and peak of the light curve the true start date and power law index cannot be well constrained. For example, \citet{auchettl2017} find a best fit power law index of $-0.92\pm0.12$ (although they bin the X-ray light curve with much larger bins) whereas \citet{2016Sci...351...62V} report an index of $-1.7$.

\section{Conclusions}\label{sec:sum}
We present radio observations of the tidal disruption event \AS\ from the Arcminute Microkelvin Imager Large Array and explore their relation to new and archival \textit{Swift} X-ray observations of the source. We find that the evolution at both wavelengths is well described by an exponential decay. We then show that the early time ($\sim200$ days) X-ray and radio emission are correlated, with a power law index of $\alpha=1.9\pm0.2$, with $L_{R}\propto L_{X}^{\alpha}$. This correlation suggests the possibility that we are observing core disk-jet coupling from \AS,\ which would contradict previous interpretations of the site of radio emission. While the X-ray emission from the source decays in a similar way throughout the observing campaign, the radio emission tails off into a plateau at $\sim250\,\mu\textrm{Jy}$. We discuss the possibility that this radio plateau is constant emission from the TDE, but conclude that it is most likely that we are now observing background radio emission from \AS's host galaxy, which we show could easily arise from supernova remnants or low level AGN activity (although we favour archival AGN activity). If the radio emission from \AS\ has continued to follow the correlation derived from the first $\sim200$ days of observations, using the most recent X-ray measurements we would expect a radio flux from the TDE of $\sim5\,\mu\textrm{Jy}$. We therefore expect that, barring a re-flaring event, radio observations of \AS\ with the AMI-LA will remain at their current level. Future X-ray and radio observations will allow us to confirm or refute our interpretation. 

\section*{Acknowledgements}

We would like to thank the staff at the Mullard Radio Astronomy Observatory for scheduling and carrying out the AMI-LA observations. This work made use of data supplied by the UK Swift Science Data Centre at the University of Leicester. JSB would like to acknowledge the support given by the Science and Technologies Facilities Council (STFC) through an STFC studentship. YCP is supported by a Trinity College JRF. The AMI telescope is supported by the European Research Council under grant ERC-2012- StG-307215 LODESTONE, the UK Science and Technology Facilities Council (STFC) and the University of Cambridge. We thank the anonymous referee for their helpful comments.

%%%%%%%%%%%%%%%%%%%%%%%%%%%%%%%%%%%%%%%%%%%%%%%%%%

%%%%%%%%%%%%%%%%%%%% REFERENCES %%%%%%%%%%%%%%%%%%%%%%

\bibliographystyle{mnras}
\bibliography{14li_paper}

\begin{thebibliography}{}
\makeatletter
\relax
\def\mn@urlcharsother{\let\do\@makeother \do\$\do\&\do\#\do\^\do\_\do\%\do\~}
\def\mn@doi{\begingroup\mn@urlcharsother \@ifnextchar [ {\mn@doi@}
  {\mn@doi@[]}}
\def\mn@doi@[#1]#2{\def\@tempa{#1}\ifx\@tempa\@empty \href
  {http://dx.doi.org/#2} {doi:#2}\else \href {http://dx.doi.org/#2} {#1}\fi
  \endgroup}
\def\mn@eprint#1#2{\mn@eprint@#1:#2::\@nil}
\def\mn@eprint@arXiv#1{\href {http://arxiv.org/abs/#1} {{\tt arXiv:#1}}}
\def\mn@eprint@dblp#1{\href {http://dblp.uni-trier.de/rec/bibtex/#1.xml}
  {dblp:#1}}
\def\mn@eprint@#1:#2:#3:#4\@nil{\def\@tempa {#1}\def\@tempb {#2}\def\@tempc
  {#3}\ifx \@tempc \@empty \let \@tempc \@tempb \let \@tempb \@tempa \fi \ifx
  \@tempb \@empty \def\@tempb {arXiv}\fi \@ifundefined
  {mn@eprint@\@tempb}{\@tempb:\@tempc}{\expandafter \expandafter \csname
  mn@eprint@\@tempb\endcsname \expandafter{\@tempc}}}

\bibitem[\protect\citeauthoryear{{Alexander}, {Berger}, {Guillochon},
  {Zauderer}  \& {Williams}}{{Alexander} et~al.}{2016}]{2016ApJ...819L..25A}
{Alexander} K.~D.,  {Berger} E.,  {Guillochon} J.,  {Zauderer} B.~A.,
  {Williams} P.~K.~G.,  2016, \mn@doi [\apjl] {10.3847/2041-8205/819/2/L25},
  \href {http://adsabs.harvard.edu/abs/2016ApJ...819L..25A} {819, L25}

\bibitem[\protect\citeauthoryear{{Alexander}, {Wieringa}, {Berger}, {Saxton}
  \& {Komossa}}{{Alexander} et~al.}{2017}]{2017ApJ...837..153}
{Alexander} K.~D.,  {Wieringa} M.~H.,  {Berger} E.,  {Saxton} R.~D.,
  {Komossa} S.,  2017, \mn@doi [\apj] {10.3847/1538-4357/aa6192}, \href
  {http://adsabs.harvard.edu/abs/2017ApJ...837..153A} {837, 153}

\bibitem[\protect\citeauthoryear{{Arcavi} et~al.,}{{Arcavi}
  et~al.}{2014}]{2014ApJ...793...38A}
{Arcavi} I.,  et~al., 2014, \mn@doi [\apj] {10.1088/0004-637X/793/1/38}, \href
  {http://adsabs.harvard.edu/abs/2014ApJ...793...38A} {793, 38}

\bibitem[\protect\citeauthoryear{{Auchettl}, {Guillochon}  \&
  {Ramirez-Ruiz}}{{Auchettl} et~al.}{2017}]{auchettl2017}
{Auchettl} K.,  {Guillochon} J.,   {Ramirez-Ruiz} E.,  2017, \mn@doi [\apj]
  {10.3847/1538-4357/aa633b}, \href
  {http://adsabs.harvard.edu/abs/2017ApJ...838..149A} {838, 149}

\bibitem[\protect\citeauthoryear{{Batejat}, {Conway}, {Hurley}, {Parra},
  {Diamond}, {Lonsdale}  \& {Lonsdale}}{{Batejat}
  et~al.}{2011}]{2011ApJ...740...95B}
{Batejat} F.,  {Conway} J.~E.,  {Hurley} R.,  {Parra} R.,  {Diamond} P.~J.,
  {Lonsdale} C.~J.,   {Lonsdale} C.~J.,  2011, \mn@doi [\apj]
  {10.1088/0004-637X/740/2/95}, \href
  {http://adsabs.harvard.edu/abs/2011ApJ...740...95B} {740, 95}

\bibitem[\protect\citeauthoryear{{Becker}, {White}  \& {Helfand}}{{Becker}
  et~al.}{1995}]{1995ApJ...450..559B}
{Becker} R.~H.,  {White} R.~L.,   {Helfand} D.~J.,  1995, \mn@doi [\apj]
  {10.1086/176166}, \href {http://adsabs.harvard.edu/abs/1995ApJ...450..559B}
  {450, 559}

\bibitem[\protect\citeauthoryear{{Bonnerot}, {Rossi}, {Lodato}  \&
  {Price}}{{Bonnerot} et~al.}{2016}]{2016MNRAS.455.2253B}
{Bonnerot} C.,  {Rossi} E.~M.,  {Lodato} G.,   {Price} D.~J.,  2016, \mn@doi
  [\mnras] {10.1093/mnras/stv2411}, \href
  {http://adsabs.harvard.edu/abs/2016MNRAS.455.2253B} {455, 2253}

\bibitem[\protect\citeauthoryear{{Brown}, {Holoien}, {Auchettl}, {Stanek},
  {Kochanek}, {Shappee}, {Prieto}  \& {Grupe}}{{Brown}
  et~al.}{2017a}]{2017MNRAS.466.4904B}
{Brown} J.~S.,  {Holoien} T.~W.-S.,  {Auchettl} K.,  {Stanek} K.~Z.,
  {Kochanek} C.~S.,  {Shappee} B.~J.,  {Prieto} J.~L.,   {Grupe} D.,  2017a,
  \mn@doi [\mnras] {10.1093/mnras/stx033}, \href
  {http://adsabs.harvard.edu/abs/2017MNRAS.466.4904B} {466, 4904}

\bibitem[\protect\citeauthoryear{{Brown} et~al.,}{{Brown}
  et~al.}{2017b}]{gbrown2017}
{Brown} G.~C.,  et~al., 2017b, \mn@doi [\mnras] {10.1093/mnras/stx2193}, \href
  {http://adsabs.harvard.edu/abs/2017MNRAS.472.4469B} {472, 4469}

\bibitem[\protect\citeauthoryear{{Burrows} et~al.,}{{Burrows}
  et~al.}{2011}]{2011Natur.476..421B}
{Burrows} D.~N.,  et~al., 2011, \mn@doi [\nat] {10.1038/nature10374}, \href
  {http://adsabs.harvard.edu/abs/2011Natur.476..421B} {476, 421}

\bibitem[\protect\citeauthoryear{{Cenko} et~al.,}{{Cenko}
  et~al.}{2012}]{2012ApJ...753...77C}
{Cenko} S.~B.,  et~al., 2012, \mn@doi [\apj] {10.1088/0004-637X/753/1/77},
  \href {http://adsabs.harvard.edu/abs/2012ApJ...753...77C} {753, 77}

\bibitem[\protect\citeauthoryear{{Condon}, {Cotton}, {Greisen}, {Yin},
  {Perley}, {Taylor}  \& {Broderick}}{{Condon}
  et~al.}{1998}]{1998AJ....115.1693C}
{Condon} J.~J.,  {Cotton} W.~D.,  {Greisen} E.~W.,  {Yin} Q.~F.,  {Perley}
  R.~A.,  {Taylor} G.~B.,   {Broderick} J.~J.,  1998, \mn@doi [\aj]
  {10.1086/300337}, \href {http://adsabs.harvard.edu/abs/1998AJ....115.1693C}
  {115, 1693}

\bibitem[\protect\citeauthoryear{{Coriat} et~al.,}{{Coriat}
  et~al.}{2011}]{coriat2011}
{Coriat} M.,  et~al., 2011, \mn@doi [\mnras]
  {10.1111/j.1365-2966.2011.18433.x}, \href
  {http://adsabs.harvard.edu/abs/2011MNRAS.414..677C} {414, 677}

\bibitem[\protect\citeauthoryear{{Davies} et~al.,}{{Davies}
  et~al.}{2009}]{davies2009}
{Davies} M.~L.,  et~al., 2009, \mn@doi [\mnras]
  {10.1111/j.1365-2966.2009.15518.x}, \href
  {http://adsabs.harvard.edu/abs/2009MNRAS.400..984D} {400, 984}

\bibitem[\protect\citeauthoryear{{Esquej}, {Saxton}, {Freyberg}, {Read},
  {Altieri}, {Sanchez-Portal}  \& {Hasinger}}{{Esquej}
  et~al.}{2007}]{esquej2007}
{Esquej} P.,  {Saxton} R.~D.,  {Freyberg} M.~J.,  {Read} A.~M.,  {Altieri} B.,
  {Sanchez-Portal} M.,   {Hasinger} G.,  2007, \mn@doi [\aap]
  {10.1051/0004-6361:20066072}, \href
  {http://adsabs.harvard.edu/abs/2007A%26A...462L..49E} {462, L49}

\bibitem[\protect\citeauthoryear{{Evans} \& {Kochanek}}{{Evans} \&
  {Kochanek}}{1989}]{1989ApJ...346L..13E}
{Evans} C.~R.,  {Kochanek} C.~S.,  1989, \mn@doi [\apjl] {10.1086/185567},
  \href {http://adsabs.harvard.edu/abs/1989ApJ...346L..13E} {346, L13}

\bibitem[\protect\citeauthoryear{{Evans} et~al.,}{{Evans}
  et~al.}{2007}]{2007A&A...469..379E}
{Evans} P.~A.,  et~al., 2007, \mn@doi [\aap] {10.1051/0004-6361:20077530},
  \href {http://adsabs.harvard.edu/abs/2007A%26A...469..379E} {469, 379}

\bibitem[\protect\citeauthoryear{{Evans} et~al.,}{{Evans}
  et~al.}{2009}]{2009MNRAS.397.1177E}
{Evans} P.~A.,  et~al., 2009, \mn@doi [\mnras]
  {10.1111/j.1365-2966.2009.14913.x}, \href
  {http://adsabs.harvard.edu/abs/2009MNRAS.397.1177E} {397, 1177}

\bibitem[\protect\citeauthoryear{{Fender} \& {Belloni}}{{Fender} \&
  {Belloni}}{2004}]{fender2004}
{Fender} R.,  {Belloni} T.,  2004, \mn@doi [\araa]
  {10.1146/annurev.astro.42.053102.134031}, \href
  {http://adsabs.harvard.edu/abs/2004ARA%26A..42..317F} {42, 317}

\bibitem[\protect\citeauthoryear{{Gehrels} et~al.,}{{Gehrels}
  et~al.}{2004}]{gehrelsSWIFT2004}
{Gehrels} N.,  et~al., 2004, \mn@doi [\apj] {10.1086/422091}, \href
  {http://adsabs.harvard.edu/abs/2004ApJ...611.1005G} {611, 1005}

\bibitem[\protect\citeauthoryear{{Gezari} et~al.,}{{Gezari}
  et~al.}{2009}]{gezariUVTDE2009}
{Gezari} S.,  et~al., 2009, \mn@doi [\apj] {10.1088/0004-637X/698/2/1367},
  \href {http://adsabs.harvard.edu/abs/2009ApJ...698.1367G} {698, 1367}

\bibitem[\protect\citeauthoryear{{Goad} et~al.,}{{Goad}
  et~al.}{2007}]{2007A&A...476.1401G}
{Goad} M.~R.,  et~al., 2007, \mn@doi [\aap] {10.1051/0004-6361:20078436}, \href
  {http://adsabs.harvard.edu/abs/2007A%26A...476.1401G} {476, 1401}

\bibitem[\protect\citeauthoryear{Guillochon \& Ramirez-Ruiz}{Guillochon \&
  Ramirez-Ruiz}{2015}]{0004-637X-809-2-166}
Guillochon J.,  Ramirez-Ruiz E.,  2015, The Astrophysical Journal, 809, 166

\bibitem[\protect\citeauthoryear{{Guillochon}, {Manukian}  \&
  {Ramirez-Ruiz}}{{Guillochon} et~al.}{2014}]{2014ApJ...783...23G}
{Guillochon} J.,  {Manukian} H.,   {Ramirez-Ruiz} E.,  2014, \mn@doi [\apj]
  {10.1088/0004-637X/783/1/23}, \href
  {http://adsabs.harvard.edu/abs/2014ApJ...783...23G} {783, 23}

\bibitem[\protect\citeauthoryear{{Hayasaki}, {Stone}  \& {Loeb}}{{Hayasaki}
  et~al.}{2013}]{2013MNRAS.434..909H}
{Hayasaki} K.,  {Stone} N.,   {Loeb} A.,  2013, \mn@doi [\mnras]
  {10.1093/mnras/stt871}, \href
  {http://adsabs.harvard.edu/abs/2013MNRAS.434..909H} {434, 909}

\bibitem[\protect\citeauthoryear{{Hickish} et~al.,}{{Hickish}
  et~al.}{2017}]{2017arXiv170704237H}
{Hickish} J.,  et~al., 2017, preprint, \href
  {http://adsabs.harvard.edu/abs/2017arXiv170704237H} {} (\mn@eprint {arXiv}
  {1707.04237})

\bibitem[\protect\citeauthoryear{{Holoien} et~al.,}{{Holoien}
  et~al.}{2016}]{holoien14li2016}
{Holoien} T.~W.-S.,  et~al., 2016, \mn@doi [\mnras] {10.1093/mnras/stv2486},
  \href {http://adsabs.harvard.edu/abs/2016MNRAS.455.2918H} {455, 2918}

\bibitem[\protect\citeauthoryear{{Irwin}, {Henriksen}, {Krause}, {Wang},
  {Wiegert}, {Murphy}, {Heald}  \& {Perlman}}{{Irwin}
  et~al.}{2015}]{2015ApJ...809..172I}
{Irwin} J.~A.,  {Henriksen} R.~N.,  {Krause} M.,  {Wang} Q.~D.,  {Wiegert} T.,
  {Murphy} E.~J.,  {Heald} G.,   {Perlman} E.,  2015, \mn@doi [\apj]
  {10.1088/0004-637X/809/2/172}, \href
  {http://adsabs.harvard.edu/abs/2015ApJ...809..172I} {809, 172}

\bibitem[\protect\citeauthoryear{{Jose} et~al.,}{{Jose}
  et~al.}{2014}]{2014ATel.6777....1J}
{Jose} J.,  et~al., 2014, The Astronomer's Telegram, \href
  {http://adsabs.harvard.edu/abs/2014ATel.6777....1J} {6777}

\bibitem[\protect\citeauthoryear{{Komossa} \& {Bade}}{{Komossa} \&
  {Bade}}{1999}]{komossa1999}
{Komossa} S.,  {Bade} N.,  1999, \aap, \href
  {http://adsabs.harvard.edu/abs/1999A%26A...343..775K} {343, 775}

\bibitem[\protect\citeauthoryear{{Krolik}, {Piran}, {Svirski}  \&
  {Cheng}}{{Krolik} et~al.}{2016}]{2016ApJ...827..127K}
{Krolik} J.,  {Piran} T.,  {Svirski} G.,   {Cheng} R.~M.,  2016, \mn@doi [\apj]
  {10.3847/0004-637X/827/2/127}, \href
  {http://adsabs.harvard.edu/abs/2016ApJ...827..127K} {827, 127}

\bibitem[\protect\citeauthoryear{{Lodato}, {King}  \& {Pringle}}{{Lodato}
  et~al.}{2009}]{2009MNRAS.392..332L}
{Lodato} G.,  {King} A.~R.,   {Pringle} J.~E.,  2009, \mn@doi [\mnras]
  {10.1111/j.1365-2966.2008.14049.x}, \href
  {http://adsabs.harvard.edu/abs/2009MNRAS.392..332L} {392, 332}

\bibitem[\protect\citeauthoryear{{McMullin}, {Waters}, {Schiebel}, {Young}  \&
  {Golap}}{{McMullin} et~al.}{2007}]{2007ASPC..376..127M}
{McMullin} J.~P.,  {Waters} B.,  {Schiebel} D.,  {Young} W.,   {Golap} K.,
  2007, in {Shaw} R.~A.,  {Hill} F.,   {Bell} D.~J.,  eds,  Astronomical
  Society of the Pacific Conference Series Vol. 376, Astronomical Data Analysis
  Software and Systems XVI. p.~127

\bibitem[\protect\citeauthoryear{{Merloni} \& {Heinz}}{{Merloni} \&
  {Heinz}}{2007}]{merloni2007}
{Merloni} A.,  {Heinz} S.,  2007, \mn@doi [\mnras]
  {10.1111/j.1365-2966.2007.12253.x}, \href
  {http://adsabs.harvard.edu/abs/2007MNRAS.381..589M} {381, 589}

\bibitem[\protect\citeauthoryear{{Merloni}, {Heinz}  \& {di Matteo}}{{Merloni}
  et~al.}{2003}]{merloni2003}
{Merloni} A.,  {Heinz} S.,   {di Matteo} T.,  2003, \mn@doi [\mnras]
  {10.1046/j.1365-2966.2003.07017.x}, \href
  {http://adsabs.harvard.edu/abs/2003MNRAS.345.1057M} {345, 1057}

\bibitem[\protect\citeauthoryear{{Miller} et~al.,}{{Miller}
  et~al.}{2015}]{2015Natur.526..542M}
{Miller} J.~M.,  et~al., 2015, \mn@doi [\nat] {10.1038/nature15708}, \href
  {http://adsabs.harvard.edu/abs/2015Natur.526..542M} {526, 542}

\bibitem[\protect\citeauthoryear{{Parra}, {Conway}, {Diamond}, {Thrall},
  {Lonsdale}, {Lonsdale}  \& {Smith}}{{Parra}
  et~al.}{2007}]{2007ApJ...659..314P}
{Parra} R.,  {Conway} J.~E.,  {Diamond} P.~J.,  {Thrall} H.,  {Lonsdale} C.~J.,
   {Lonsdale} C.~J.,   {Smith} H.~E.,  2007, \mn@doi [\apj] {10.1086/511813},
  \href {http://adsabs.harvard.edu/abs/2007ApJ...659..314P} {659, 314}

\bibitem[\protect\citeauthoryear{{Pasham} \& {van Velzen}}{{Pasham} \& {van
  Velzen}}{2017}]{pasham2017}
{Pasham} D.~R.,  {van Velzen} S.,  2017, preprint, \href
  {http://adsabs.harvard.edu/abs/2017arXiv170902882P} {} (\mn@eprint {arXiv}
  {1709.02882})

\bibitem[\protect\citeauthoryear{{Pasham}, {Cenko}, {Sadowski}, {Guillochon},
  {Stone}, {van Velzen}  \& {Cannizzo}}{{Pasham}
  et~al.}{2017}]{2017ApJ...837L..30P}
{Pasham} D.~R.,  {Cenko} S.~B.,  {Sadowski} A.,  {Guillochon} J.,  {Stone}
  N.~C.,  {van Velzen} S.,   {Cannizzo} J.~K.,  2017, \mn@doi [\apjl]
  {10.3847/2041-8213/aa6003}, \href
  {http://adsabs.harvard.edu/abs/2017ApJ...837L..30P} {837, L30}

\bibitem[\protect\citeauthoryear{{Phinney}}{{Phinney}}{1989}]{phinney1989}
{Phinney} E.~S.,  1989, in {Morris} M.,  ed.,  IAU Symposium Vol. 136, The
  Center of the Galaxy. p.~543

\bibitem[\protect\citeauthoryear{{Piran}, {Svirski}, {Krolik}, {Cheng}  \&
  {Shiokawa}}{{Piran} et~al.}{2015}]{piranSELF2015}
{Piran} T.,  {Svirski} G.,  {Krolik} J.,  {Cheng} R.~M.,   {Shiokawa} H.,
  2015, \mn@doi [\apj] {10.1088/0004-637X/806/2/164}, \href
  {http://adsabs.harvard.edu/abs/2015ApJ...806..164P} {806, 164}

\bibitem[\protect\citeauthoryear{Prieto et~al.,}{Prieto
  et~al.}{2016}]{2041-8205-830-2-L32}
Prieto J.~L.,  et~al., 2016, The Astrophysical Journal Letters, 830, L32

\bibitem[\protect\citeauthoryear{{Rees}}{{Rees}}{1988}]{1988Natur.333..523R}
{Rees} M.~J.,  1988, \mn@doi [\nat] {10.1038/333523a0}, \href
  {http://adsabs.harvard.edu/abs/1988Natur.333..523R} {333, 523}

\bibitem[\protect\citeauthoryear{{Romero-Ca{\~n}izales}, {Prieto}, {Chen},
  {Kochanek}, {Dong}, {Holoien}, {Stanek}  \& {Liu}}{{Romero-Ca{\~n}izales}
  et~al.}{2016}]{2016ApJ...832L..10R}
{Romero-Ca{\~n}izales} C.,  {Prieto} J.~L.,  {Chen} X.,  {Kochanek} C.~S.,
  {Dong} S.,  {Holoien} T.~W.-S.,  {Stanek} K.~Z.,   {Liu} F.,  2016, \mn@doi
  [\apjl] {10.3847/2041-8205/832/1/L10}, \href
  {http://adsabs.harvard.edu/abs/2016ApJ...832L..10R} {832, L10}

\bibitem[\protect\citeauthoryear{{Smith}, {Lonsdale}, {Lonsdale}  \&
  {Diamond}}{{Smith} et~al.}{1998}]{1998ApJ...493L..17S}
{Smith} H.~E.,  {Lonsdale} C.~J.,  {Lonsdale} C.~J.,   {Diamond} P.~J.,  1998,
  \mn@doi [\apjl] {10.1086/311122}, \href
  {http://adsabs.harvard.edu/abs/1998ApJ...493L..17S} {493, L17}

\bibitem[\protect\citeauthoryear{{Swinbank} et~al.,}{{Swinbank}
  et~al.}{2015}]{2015A&C....11...25S}
{Swinbank} J.~D.,  et~al., 2015, \mn@doi [Astronomy and Computing]
  {10.1016/j.ascom.2015.03.002}, \href
  {http://adsabs.harvard.edu/abs/2015A%26C....11...25S} {11, 25}

\bibitem[\protect\citeauthoryear{{Voges} et~al.,}{{Voges}
  et~al.}{1999}]{1999A&A...349..389V}
{Voges} W.,  et~al., 1999, \aap, \href
  {http://adsabs.harvard.edu/abs/1999A%26A...349..389V} {349, 389}

\bibitem[\protect\citeauthoryear{{Zauderer} et~al.,}{{Zauderer}
  et~al.}{2011}]{2011Natur.476..425Z}
{Zauderer} B.~A.,  et~al., 2011, \mn@doi [\nat] {10.1038/nature10366}, \href
  {http://adsabs.harvard.edu/abs/2011Natur.476..425Z} {476, 425}

\bibitem[\protect\citeauthoryear{{Zwart} et~al.,}{{Zwart}
  et~al.}{2008}]{2008MNRAS.391.1545Z}
{Zwart} J.~T.~L.,  et~al., 2008, \mn@doi [\mnras]
  {10.1111/j.1365-2966.2008.13953.x}, \href
  {http://ads.ari.uni-heidelberg.de/abs/2008MNRAS.391.1545Z} {391, 1545}

\bibitem[\protect\citeauthoryear{{van Velzen} et~al.,}{{van Velzen}
  et~al.}{2011}]{vanvelzenOPTTDE2011}
{van Velzen} S.,  et~al., 2011, \mn@doi [\apj] {10.1088/0004-637X/741/2/73},
  \href {http://adsabs.harvard.edu/abs/2011ApJ...741...73V} {741, 73}

\bibitem[\protect\citeauthoryear{{van Velzen} et~al.,}{{van Velzen}
  et~al.}{2016}]{2016Sci...351...62V}
{van Velzen} S.,  et~al., 2016, \mn@doi [Science] {10.1126/science.aad1182},
  \href {http://adsabs.harvard.edu/abs/2016Sci...351...62V} {351, 62}

\makeatother
\end{thebibliography}

%%%%%%%%%%%%%%%%%%%%%%%%%%%%%%%%%%%%%%%%%%%%%%%%%%

%%%%%%%%%%%%%%%%% APPENDICES %%%%%%%%%%%%%%%%%%%%%

\appendix

\section{Observations}

\subsection{Radio Observations}
\onecolumn
\begin{longtable}{cccc}
\caption{AMI-LA radio observations of \AS.}\\\hline
Date (MJD) & Frequency & Flux density & Flux density error\\ 
(MJD) & (GHz) & ($\mu\textrm{Jy}$) & ($\mu\textrm{Jy}$)\\
\hline\hline
\endfirsthead
\caption{AMI-LA radio observations of \AS.}\\\hline
Date & Frequency & Flux density & Flux density error\\ 
(MJD) & (GHz) & ($\mu\textrm{Jy}$) & ($\mu\textrm{Jy}$)\\\hline\hline
\endhead
\hline
\endfoot
57014.078 & 15.7         & 1865                       & 83                            \\
57015.170 & 15.7         & 1871                       & 141                           \\
57017.096 & 15.7         & 1969                       & 67                            \\
57021.111 & 15.7         & 2096                       & 84                            \\
57039.088 & 15.7         & 1629                       & 78                            \\
57061.053 & 15.7         & 969                        & 150                           \\
57065.995 & 15.7         & 870                        & 103                           \\
57068.089 & 15.7         & 879                        & 72                            \\
57069.970 & 15.7         & 1291                       & 209                           \\
57072.982 & 15.7         & 836                        & 286                           \\
57077.055 & 15.7         & 860                        & 86                            \\
57079.057 & 15.7         & 652                        & 93                            \\
57082.105 & 15.7         & 654                        & 66                            \\
57085.088 & 15.7         & 781                        & 61                            \\
57090.081 & 15.7         & 894                        & 121                           \\
57094.062 & 15.7         & 789                        & 90                            \\
57098.952 & 15.7         & 761                        & 82                            \\
57109.041 & 15.7         & 779                        & 127                           \\
57113.984 & 15.7         & 826                        & 367                           \\
57118.978 & 15.7         & 740                        & 103                           \\
57122.941 & 15.7         & 814                        & 104                           \\
57125.982 & 15.7         & 708                        & 87                            \\
57127.975 & 15.7         & 717                        & 76                            \\
57133.958 & 15.7         & 715                        & 117                           \\
57138.941 & 15.7         & 525                        & 73                            \\
57142.951 & 15.7         & 560                        & 79                            \\
57145.921 & 15.7         & 497                        & 99                            \\
57150.749 & 15.7         & 617                        & 191                           \\
57153.735 & 15.7         & 503                        & 177                           \\
57374.323 & 15.5         & 245                        & 57                            \\
57376.128 & 15.5         & 224                        & 65                            \\
57379.112 & 15.5         & 307                        & 65                            \\
57381.291 & 15.5         & 336                        & 70                            \\
57386.186 & 15.5         & 360                        & 93                            \\
57388.098 & 15.5         & 305                        & 68                            \\
57390.092 & 15.5         & 265                        & 88                            \\
57392.087 & 15.5         & 132                        & 89                            \\
57394.081 & 15.5         & 298                        & 114                           \\
57396.066 & 15.5         & 257                        & 93                            \\
57399.243 & 15.5         & 289                        & 60                            \\
57401.211 & 15.5         & 216                        & 59                            \\
57404.097 & 15.5         & 231                        & 69                            \\
57406.156 & 15.5         & 257                        & 61                            \\
57409.182 & 15.5         & 262                        & 63                            \\
57413.019 & 15.5         & 274                        & 68                            \\
57419.077 & 15.5         & 194                        & 63                            \\
57421.028 & 15.5         & 255                        & 66                            \\
57422.981 & 15.5         & 310                        & 71                            \\
57428.013 & 15.5         & 191                        & 72                            \\
57431.133 & 15.5         & 253                        & 51                            \\
57434.038 & 15.5         & 333                        & 87                            \\
57438.027 & 15.5         & 278                        & 63                            \\
57441.950 & 15.5         & 286                        & 87                            \\
57446.040 & 15.5         & 278                        & 46                            \\
57461.083 & 15.5         & 201                        & 49                            \\
57464.874 & 15.5         & 154                        & 61                            \\
57467.897 & 15.5         & 213                        & 60                            \\
57482.937 & 15.5         & 219                        & 86                            \\
57496.982 & 15.5         & 211                        & 57                            \\
57500.989 & 15.5         & 169                        & 72                            \\
57508.040 & 15.5         & 200                        & 83                            \\
57534.795 & 15.5         & 226                        & 58                            \\
57538.815 & 15.5         & 256                        & 76                            \\
57575.815 & 15.5         & 305                        & 96                            \\
57632.565 & 15.5         & 143                        & 99                            \\
57634.542 & 15.5         & 293                        & 108                           \\
57640.501 & 15.5         & 202                        & 78                            \\
57649.483 & 15.5         & 187                        & 89                            \\
57653.480 & 15.5         & 274                        & 72                            \\
57661.399 & 15.5         & 265                        & 108                           \\
57668.399 & 15.5         & 254                        & 105                           \\
57675.482 & 15.5         & 230                        & 67                            \\
57678.358 & 15.5         & 242                        & 81                            \\
57688.384 & 15.5         & 211                        & 69                            \\
57695.340 & 15.5         & 213                        & 64                            \\
57696.316 & 15.5         & 161                        & 78                            \\
57700.305 & 15.5         & 249                        & 77                            \\
57710.302 & 15.5         & 254                        & 53                            \\
57717.262 & 15.5         & 301                        & 58                            \\
57720.272 & 15.5         & 254                        & 61                            \\
57721.227 & 15.5         & 289                        & 70                            \\
57731.224 & 15.5         & 258                        & 90                            \\
57737.204 & 15.5         & 216                        & 61                            \\
57745.226 & 15.5         & 214                        & 77                            \\
57749.236 & 15.5         & 196                        & 62                            \\
57752.184 & 15.5         & 277                        & 74                            \\
57761.183 & 15.5         & 216                        & 70                            \\
57765.172 & 15.5         & 248                        & 68                            \\
57769.141 & 15.5         & 199                        & 95                            \\
57777.143 & 15.5         & 165                        & 64                            \\
57783.100 & 15.5         & 195                        & 130                           \\
57788.087 & 15.5         & 210                        & 52                            \\
57794.050 & 15.5         & 266                        & 78                            \\
57799.122 & 15.5         & 218                        & 89                            \\
57809.033 & 15.5         & 209                        & 58                            \\
57814.037 & 15.5         & 246                        & 57                            \\
57820.918 & 15.5         & 231                        & 54                            \\
57829.971 & 15.5         & 199                        & 77                            \\
57832.963 & 15.5         & 244                        & 57                            \\
57834.958 & 15.5         & 336                        & 98                            \\
57839.945 & 15.5         & 265                        & 74                            \\
57888.911 & 15.5         & 188                        & 119                           \\
57920.723 & 15.5         & 193                        & 85                            \\
57949.661 & 15.5         & 261                        & 91
\end{longtable}

\subsection{X-ray Observations}
\onecolumn
\begin{longtable}{ccccc}
\caption{\textit{Swift} $0.3$-$10\,\textrm{keV}$ observations of \AS.}\\\hline
Date & Count rate & Count rate error & Flux & Flux error\\
(MJD) & ($\textrm{counts}\,\textrm{s}^{-1}$) & ($\textrm{counts}\,\textrm{s}^{-1}$) & ($10^{12}\,\textrm{erg}\,\textrm{s}^{-1}\,\textrm{cm}^{-2}$) & ($10^{12}\,\textrm{erg}\,\textrm{s}^{-1}\,\textrm{cm}^{-2}$)\\\hline\hline
\endfirsthead
\caption{\textit{Swift} $0.3$-$10\,\textrm{keV}$ observations of \AS.}\\\hline
Date & Count rate & Count rate error & Flux & Flux error\\
(MJD) & ($\textrm{counts}\,\textrm{s}^{-1}$) & ($\textrm{counts}\,\textrm{s}^{-1}$) & ($10^{12}\,\textrm{erg}\,\textrm{s}^{-1}\,\textrm{cm}^{-2}$) & ($10^{12}\,\textrm{erg}\,\textrm{s}^{-1}\,\textrm{cm}^{-2}$)\\\hline\hline
\endhead
\hline
\endfoot
56998.259  & 0.410         & 0.016               & 18.8    & 0.7        \\
57001.637  & 0.397         & 0.017               & 18.2    & 0.8        \\
57004.197  & 0.429         & 0.024               & 19.7    & 1.0        \\
57007.296  & 0.454         & 0.017               & 20.8    & 0.8        \\
57010.835  & 0.524         & 0.018               & 24.0    & 0.8        \\
57013.099  & 0.448         & 0.016               & 20.6    & 0.7        \\
57016.092  & 0.435         & 0.019               & 20.0    & 0.9        \\
57019.576  & 0.390         & 0.015               & 17.9    & 0.7        \\
57022.745  & 0.377         & 0.016               & 17.3    & 0.7        \\
57029.582  & 0.494         & 0.015               & 22.7    & 0.7        \\
57033.144  & 0.474         & 0.023               & 21.7    & 1.0        \\
57036.110  & 0.456         & 0.021               & 20.9    & 1.0        \\
57039.234  & 0.401         & 0.016               & 18.4    & 0.7        \\
57042.296  & 0.394         & 0.018               & 18.1    & 0.8        \\
57045.624  & 0.388         & 0.015               & 17.8    & 0.7        \\
57048.821  & 0.400         & 0.016               & 18.4    & 0.7        \\
57051.534  & 0.363         & 0.017               & 16.7    & 0.8        \\
57054.139  & 0.337         & 0.012               & 15.5    & 0.6        \\
57057.560  & 0.329         & 0.013               & 15.1    & 0.6        \\
57060.165  & 0.337         & 0.012               & 15.5    & 0.6        \\
57065.849  & 0.243         & 0.014               & 11.2    & 0.6        \\
57068.779  & 0.269         & 0.015               & 12.3    & 0.7        \\
57071.737  & 0.278         & 0.012               & 12.8    & 0.6        \\
57074.910  & 0.270         & 0.011               & 12.4    & 0.5        \\
57077.633  & 0.231         & 0.013               & 10.6    & 0.6        \\
57081.189  & 0.256         & 0.010               & 11.8    & 0.5        \\
57086.917  & 0.299         & 0.013               & 13.7    & 0.6        \\
57089.381  & 0.327         & 0.013               & 15.0    & 0.6        \\
57099.421  & 0.286         & 0.016               & 13.1    & 0.7        \\
57102.655  & 0.276         & 0.012               & 12.7    & 0.6        \\
57105.311  & 0.175         & 0.019               & 8.0     & 0.9        \\
57109.208  & 0.256         & 0.012               & 11.8    & 0.6        \\
57111.933  & 0.242         & 0.012               & 11.1    & 0.6        \\
57114.128  & 0.220         & 0.014               & 10.1    & 0.6        \\
57117.720  & 0.199         & 0.011               & 9.1     & 0.5        \\
57120.317  & 0.193         & 0.011               & 8.9     & 0.5        \\
57123.578  & 0.195         & 0.010               & 9.0     & 0.5        \\
57126.243  & 0.178         & 0.009               & 8.2     & 0.4        \\
57129.400  & 0.191         & 0.012               & 8.8     & 0.6        \\
57132.560  & 0.162         & 0.010               & 7.4     & 0.5        \\
57136.560  & 0.156         & 0.012               & 7.2     & 0.6        \\
57139.347  & 0.190         & 0.011               & 8.7     & 0.5        \\
57147.597  & 0.187         & 0.010               & 8.6     & 0.5        \\
57150.256  & 0.189         & 0.009               & 8.7     & 0.4        \\
57153.486  & 0.183         & 0.009               & 8.4     & 0.4        \\
57156.647  & 0.157         & 0.010               & 7.2     & 0.5        \\
57173.106  & 0.161         & 0.010               & 7.4     & 0.5        \\
57176.134  & 0.155         & 0.010               & 7.1     & 0.5        \\
57179.194  & 0.152         & 0.016               & 7.0     & 0.7        \\
57182.465  & 0.155         & 0.013               & 7.1     & 0.6        \\
57186.055  & 0.148         & 0.010               & 6.8     & 0.5        \\
57188.549  & 0.189         & 0.039               & 8.7     & 2.0        \\
57191.845  & 0.154         & 0.009               & 7.1     & 0.4        \\
57195.195  & 0.156         & 0.009               & 7.2     & 0.4        \\
57200.382  & 0.131         & 0.011               & 6.0     & 0.5        \\
57203.807  & 0.120         & 0.009               & 5.5     & 0.4        \\
57226.617  & 0.090         & 0.007               & 4.1     & 0.3        \\
57230.379  & 0.103         & 0.011               & 4.7     & 0.5        \\
57236.470  & 0.096         & 0.007               & 4.4     & 0.3        \\
57238.860  & 0.111         & 0.008               & 5.1     & 0.4        \\
57242.120  & 0.104         & 0.007               & 4.8     & 0.3        \\
57246.907  & 0.075         & 0.006               & 3.5     & 0.3        \\
57340.745  & 0.079         & 0.006               & 3.6     & 0.3        \\
57351.843  & 0.067         & 0.007               & 3.1     & 0.3        \\
57354.698  & 0.063         & 0.006               & 2.9     & 0.3        \\
57357.365  & 0.065         & 0.006               & 3.0     & 0.3        \\
57360.258  & 0.080         & 0.007               & 3.7     & 0.3        \\
57363.951  & 0.089         & 0.006               & 4.1     & 0.3        \\
57366.940  & 0.084         & 0.006               & 3.9     & 0.3        \\
57369.900  & 0.071         & 0.006               & 3.3     & 0.3        \\
57372.255  & 0.063         & 0.007               & 2.9     & 0.3        \\
57375.579  & 0.088         & 0.009               & 4.0     & 0.4        \\
57378.411  & 0.078         & 0.007               & 3.6     & 0.3        \\
57383.203  & 0.074         & 0.023               & 3.4     & 1.0        \\
57411.684  & 0.056         & 0.006               & 2.6     & 0.3        \\
57417.730  & 0.084         & 0.008               & 3.9     & 0.4        \\
57423.806  & 0.048         & 0.009               & 2.2     & 0.4        \\
57426.466  & 0.032         & 0.011               & 1.5     & 0.5        \\
57427.758  & 0.050         & 0.006               & 2.3     & 0.3        \\
57429.669  & 0.050         & 0.012               & 2.3     & 0.6        \\
57433.319  & 0.044         & 0.006               & 2.0     & 0.3        \\
57435.743  & 0.048         & 0.005               & 2.2     & 0.2        \\
57519.819  & 0.028         & 0.004               & 1.3     & 0.2        \\
57522.607  & 0.021         & 0.003               & 1.0     & 0.1        \\
57526.827  & 0.020         & 0.004               & 0.9     & 0.2        \\
57542.686  & 0.022         & 0.004               & 1.0     & 0.2        \\
57545.455  & 0.021         & 0.009               & 1.0     & 0.4        \\
57546.401  & 0.024         & 0.005               & 1.1     & 0.2        \\
57550.131  & 0.032         & 0.005               & 1.5     & 0.2        \\
57554.383  & 0.013         & 0.003               & 0.6     & 0.1        \\
57718.041  & 0.022         & 0.003               & 1.0     & 0.1        \\
57820.434  & 0.008         & 0.003               & 0.4     & 0.1        \\
57821.966  & 0.015         & 0.003               & 0.7     & 0.1        \\
57826.917  & 0.013         & 0.003               & 0.6     & 0.1        \\
57828.568  & 0.014         & 0.005               & 0.7     & 0.2       \\\hline
\end{longtable}

%%%%%%%%%%%%%%%%%%%%%%%%%%%%%%%%%%%%%%%%%%%%%%%%%%

% Don't change these lines
\bsp	% typesetting comment
\label{lastpage}
\end{document}